\documentclass{pasj00}
\draft

\begin{document}
\SetRunningHead{K. Ogura et al.}{Age Sequence in Bright-Rim Aggregates}
\Received{2006/10/10}%{yyyy/mm/dd}
\Accepted{2006/xx/xx}%{yyyy/mm/dd}

\title{Age Sequence in Small Clusters Associated with Bright-Rimmed Clouds}

%%% begin:list of authors
\author{Katsuo {\textsc Ogura}, \altaffilmark{1}
              Neelam {\textsc Chauhan},  \altaffilmark{2}
              Anil K. {\textsc Pandey}, \altaffilmark{2}
              Bhuwan C. {\textsc Bhatt}, \altaffilmark{3}
              Devendra {\textsc Ojha}, \altaffilmark{4} \\
              and 
             Yoichi {\textsc Itoh}, \altaffilmark{5}
             }
%\thanks{Example: Present Address is xxxxxxxxxx}}
\altaffiltext{1}{Kokugakuin University, Higashi, Shibuya-ku, Tokyo 150-8440}
\email{ogura@kokugakuin.ac.jp}

\altaffiltext{2}{Aryabhatta Research Institute of Observational Sciences, Manora Peak, Naini Tal - 263 129, India}%\email{ccccc@xxx.xxx.xx.xx}

\altaffiltext{3}{CREST, Indian Institute of Astrophysics, Shidlaghatta Road, Hosakote - 562 114, India}%\email{ccccc@xxx.xxx.xx.xx}

\altaffiltext{4}{Tata Institute of Fundamental Research, Homi Bhabha Road, Colaba, Mumbai - 400 005, India}%\email{ccccc@xxx.xxx.xx.xx}

\altaffiltext{5}{Graduate School of Science and Technology, Kobe University, 1-1 Rokkodai, Nada, Kobe, Hyogo 657-8501}%\email{ccccc@xxx.xxx.xx.xx}

%%% end:list of authors

%%% Please use the following style in case that sorting by 
%%% affilation is impossible. 
%
% \author{%
%   D-Firstname \textsc{D-Familyname}\altaffilmark{1}
%   E-Firstname \textsc{E-Familyname}\altaffilmark{1,2}
%   and
%   F-Firstname \textsc{F-Familyname}\altaffilmark{2}}
% \altaffiltext{1}{Address of Institute}
% \email{ddddd@xxx.xxx.xx.xx}
% \email{eeeee@xxx.xxx.xx.xx}
% \altaffiltext{2}{Address of Institute}

%% `\KeyWords{}' always has to be placed before `\maketitle'.

 %Do NOT move this preamble from here!

\maketitle

\begin{abstract}

Bright-rimmed clouds (BRCs) found in H\emissiontype{II} regions are probable sites of triggered star formation due to compression by ionization/shock fronts, and it is hypothesized that star formation proceeds from the exciting star(s) side outward of the H\emissiontype{II} region ("{\it small-scale sequential star formation\/}"). In order to quantitatively testify this hypothesis we undertook {\it BVI$_{c}$\/} photometry of four BRC aggregates. The amounts of interstellar extinction and reddening for each star have been estimated by using the {\it JHK$_{s}$\/} photometry.  Then we constructed reddening-corrected {\it V}/{\it V-I$_{c}$\/} color-magnitude diagrams, where the age of each star has been derived. All the stars turned out to be a few tenths to a few Myr old. Although the scatters are large and the numbers of the sample stars are small, we found a clear trend that the stars inside or in the immediate vicinity of the bright rim are younger than those outside it in all the four aggregates, confirming the hypothesis in question. 

\end{abstract}

\section{Introduction}

\noindent 

Bright-rimmed clouds (BRCs), or cometary globules, correspond to relatively dense clumps in a giant molecular cloud left unionized in the course of the expansion of the H\emissiontype{II} region. So they are a sort of remnants of star formation activity in giant molecular clouds. But, at the same time, they are current sites of star formation. Theoretically, triggered star formation caused by the compression of the gas due to  shock is expected to take place in such a pre-existing dense clump in an H\emissiontype{II} region. This phenomenon is called radiation-driven implosion (RDI) and detailed model calculations were carried out by several authors (e.g., \cite{ber89}, \cite{lef94}, \cite{mia06}). Many molecular line and radio continuum observations of BRCs (\cite{lef95}, \cite{lef97}, \cite{dev02}, \cite{tho04a}, \cite{tho04b}, \cite{tho04c}, \cite{mor04}, \cite{urq06}) show that the overall morphology and some physical properties of them are consistent with those of the RDI models. 

Actually, many BRCs are associated with signposts of recent and ongoing star formation such as Herbig-Haro objects and IRAS point sources of low temperature that meet the YSO criteria of, e.g.,  \citet{car00}. \citet{sug91} and \citet{sug94} compiled catalogs (the so-called SFO Catalog) of altogether 90 BRCs associated with such IRAS point sources for the northern and southern skies, respectively. Subsequent near-IR ({\it JHK\/}) imaging observations of many of the BRCs in the above catalogs revealed that an elongated, small cluster or aggregate of YSOs are often associated with them (\cite{sug95}). Interestingly, "redder" stars tend to be located inside the BRC and closer to the IRAS source that corresponds to the reddest source and lies at the innermost end of the aggregate, and "bluer" stars tend to lie outside the cloud on the side of  the O star(s) exciting the H\emissiontype{II} region. These results led them to advocate the "small-scale sequential star formation (hereafter, {\it S\/$^4$F\/}) hypothesis" that a BRC undergoes a few or several star formation events due to RDI always at its head part, which is located closer to the exciting stars(s) first, but then recedes outward in the H\emissiontype{II} region as the the star formation proceeds, leaving groups of stars aligned in an age sequence.

With the aim to strengthen the {\it S\/$^4$F\/} hypothesis \citet{ogu02} made grism surveys of 24 BRCs for H$\alpha$ emission stars. Altogether 460 H$\alpha$ emission stars and 12 Herbig-Haro objects have been detected in their vicinities. Presumably these H$\alpha$ emission stars are mostly T Tauri stars (TTSs), although practically no follow-up observations have yet been done for them. These objects are found concentrated toward the head or just outside of BRCs on the side of the exciting star(s). One may suspect that there might be just as many such stars farther inside the BRCs, but concealed by the higher extinction. However this does not appear likely, because, as we see in Sect. 4, the {\it A$_{V}$\/} values of the H$\alpha$ emission stars inside, e.g., BRC 14 are not much larger than those outside it. In addition in figures 1a to 1z of \citet{ogu02} we do not find many faint (H$\alpha$ emission) stars inside BRCs. So we think that the observed concentration of H$\alpha$ emitters toward the heads of the BRCs supports the {\it S\/$^4$F\/} hypothesis, although it may be exaggerated to some extent by the opacity effect. (Here we also add that as for the very rich AFGL 4029 IR cluster embedded in somewhat eastern part of BRC 14 (e.g., \cite{deh97}) we believe that it formed spontaneously, not as a result of RDI; it is located too far away from the head of the BRC 14 for the effects of RDI to propagate.) Recently \citet{mat06} analyzed their deep {\it JHK\/}photometry of the BRC 14 region and showed that three indicators of star formation, i.e., the  fraction of YSO candidates, the amount of extinction of all sources, and near-IR excesses of the YSO candidates, all showed a clear sequence from outside to inside the bright rim. This result further strengthens the {\it S\/$^4$F\/} hypothesis. 

The best way to quantitatively testify the hypothesis is to estimate of the ages of the aggregate members and to compare them between different regions with respect to the bright rim. So we undertook {\it BVI$_{c}$\/} photometry of BRC aggregates. The problems associated with this attempt, in contrast to the case of ordinary open clusters, are membership selection and reddening correction. We  overcame these difficulties as described in Sec. 3, and constructed reddening-corrected  color-magnitude diagrams (CMDs) for each BRC aggregate. The age of each star has been determined on the {\it V}$_{o}$/({\it V-I$_{c}$\/})$_{o}$  CMDs, and its spatial distribution has been examined. The result seems to confirm the  {\it S\/$^4$F\/} hypothesis.

%\newpage

\begin{table}
  \caption{Log of observations}\label{tab:first}
  \begin{center}
    \begin{tabular}{lll}
     \hline\hline
 BRC & Date of Obs. & Exposure Times (sec) \\
  ~ & ~ & $\times$ No. of Frames \\
  \hline 
      11NE & 05/9/27 &  {\it I$_{c}$\/}: 60$\times$4  ~{\it V\/}: 120$\times$4  ~{\it B\/}: 300$\times$4 \\
      12  &   05/9/27 &   {\it I$_{c}$\/}: 60$\times$4  ~{\it V\/}: 100$\times$4  ~{\it B\/}: 240$\times$4 \\
      14  &   05/9/27 &   {\it I$_{c}$\/}: 50$\times$4  ~{\it V\/}: 120$\times$4  ~{\it B\/}: 300$\times$4 \\
      37  &   05/9/25 &   {\it I$_{c}$\/}: 60$\times$6  ~{\it V\/}: 200$\times$6  ~{\it B\/}: 300$\times$6 \\
     \hline
     \end{tabular}
  \end{center}
\end{table}

%%%%%%%%%%%%%%%%%%%%%%%%%%%%%%%%%%%%%%%

\section{Observations and Data Reduction}

The imaging observations of the four BRCs in the {\it BVI$_{c}$\/} bands were carried out on the 2.0-m Himalayan Chandra Telescope (HCT) of the Indian Astronomical Observatory. HCT is located at Mt. Saraswati (altitude 4500m above see level) in the Himalayan region, but is remotely operated from CREST, Hosakote near Bangalore, via a satellite link. The instrument used is HFOSC in the imaging mode. The details of the site, HCT and HFOSC can be found at the HCT website\footnote{{\tt http://www.crest.ernet.in/iia/iao/iao.html}}. The observations were made on 25 and 27 Sep, 2005 and a log of the observations is given in table 1. The sky conditions at the time of the observations were satisfactory with the seeing size of about $1.5''$.

The data reduction was carried out at Aryabhatta Research Institute of Observational Sciences (ARIES), India. The initial processing of the data frames was made using the IRAF data reduction package. The photometric measurements were done using DAOPHOT II (\cite{ste87}). The point spread function (psf) was obtained for each frame using several uncontaminated stars. The photometric accuracies depend on the brightness of the stars, and the typical DAOPHOT errors of our target stars ({\it V\/} $\sim$ 18 mag) are: 0.003 mag in {\it I$_{c}$\/}, 0.005 mag in {\it V\/}, and 0.007 mag in {\it B\/} for BRCs 11NE, 12 and 37. However they are a little bit larger for BRC 14: 0.005, 0.005 and 0.007, respectively. The standard deviation of the residuals of the transformed {\it V\/} magnitude and {\it B-V\/} and {\it V-I$_{c}$\/} colors of the standard stars were 0.010, 0.005 and 0.012, respectively. Near the limiting magnitude of {\it V\/} $\sim$ 20.5, the DAOPHOT errors increase to $\sim$0.03 in all color bands.

\section{Star Selection and Reddening Correction}

BRC aggregates are very loose and composed of a small number of stars. In addition their Galactic latitudes are low, so there can be many field stars, although some of them may be associated with the H\emissiontype{II} region to which the BRC belongs. Therefore we selected the targets from such BRCs that the associated aggregate contains relatively many stars. BRCs 11NE, 12 and 14 belong to the H\emissiontype{II} region IC 1848 and BRC 37 to IC 1396. For each BRC the H$\alpha$ emission stars of \citet{ogu02} have been selected as the member stars. In the case of BRC 11NE and BRC 37, however, only a few H$\alpha$ stars are available. But, as one sees in figures 1f and 1p of \citet{ogu02}, a relatively clear aggregate of stars can be noticed that appears to be associated with the BRCs; these aggregates show up more prominently in the 2MASS images. We, therefore, have added the stars having 2MASS counterparts from the aggregates, and they are indicated with "A" ({\it additional stars\/}) in the finding charts, figure 1. In  BRCs 12 and 14 we deal only with the H$\alpha$ stars of \citet{ogu02}, so see the finding charts therein. 

Another difficulty is the extinction correction. The target stars must have different amounts of extinction caused by the dust in the BRC and/or H\emissiontype{II} regions. Moreover, they are thought to be pre-main sequence stars and so have peculiar color indices, in particular, {\it U-B\/} ("UV excess"). Therefore the ordinary technique of {\it U-B}/{\it B-V\/} two-color diagram (2CD) to estimate the amount of reddening in open clusters cannot be used. {\it B-V\/} and {\it V-I$_{c}$\/} are less peculiar in TTSs, but a new problem emerges: in the {\it B-V}/{\it V-I$_{c}$\/} 2CD the intrinsic line is nearly parallel to the reddening vector in the part for late-type stars, so this 2CD is practically not usable to estimate the amount of reddening.

In order to overcome this difficulty we used the {\it J-H/H-K$_{s}$\/} 2CD, where the  {\it A$_{V}$\/} value for each star has been measured by tracing back to the intrinsic lines along the reddening vector found in \citet{mey97}. We adopted the intrinsic line of TTSs given by \citet{mey97} and that of dwarfs of  \citet{bes88}. The former is used for H$\alpha$ stars. The latter, used for non-emission stars, have been converted into the CIT {\it JHK$_{s}$\/} system according to the equation given in \citet{bes88}. For BRCs 11NE, 12 and 37 the {\it JHK$_{s}$\/} data have been taken from the 2MASS catalog and converted into the CIT {\it JHK$_{s}$\/} system according to the relation given in the 2MASS web site\footnote{{\tt  http://www.astro.caltech.edu/\%7Ejmc/2mass/v3/transformations}}. Only stars with the photometric qualities 'AAA' have been adopted. For BRC 14 we have adopted the {\it JHK$_{s}$\/} data used in \citet{mat06}, which were obtained with the infrared camera SIRIUS on the University of Hawaii 2.2-m telescope atop Mauna Kea. The identifications of the target stars between the HCT data and the 2MASS catalog or the SIRIUS data have been made on the basis of the J2000.0 coordinates within an accuracy of $2''$. From the  {\it A$_{V}$\/} values thus obtained we have calculated the color excesses {\it E$_{V-Ic}$\/} and  {\it E$_{B-V}$\/} according to the ratio given by \citet{mun96}. 

%\newpage

\begin{figure}
  \begin{center}
    \FigureFile(160mm,80mm){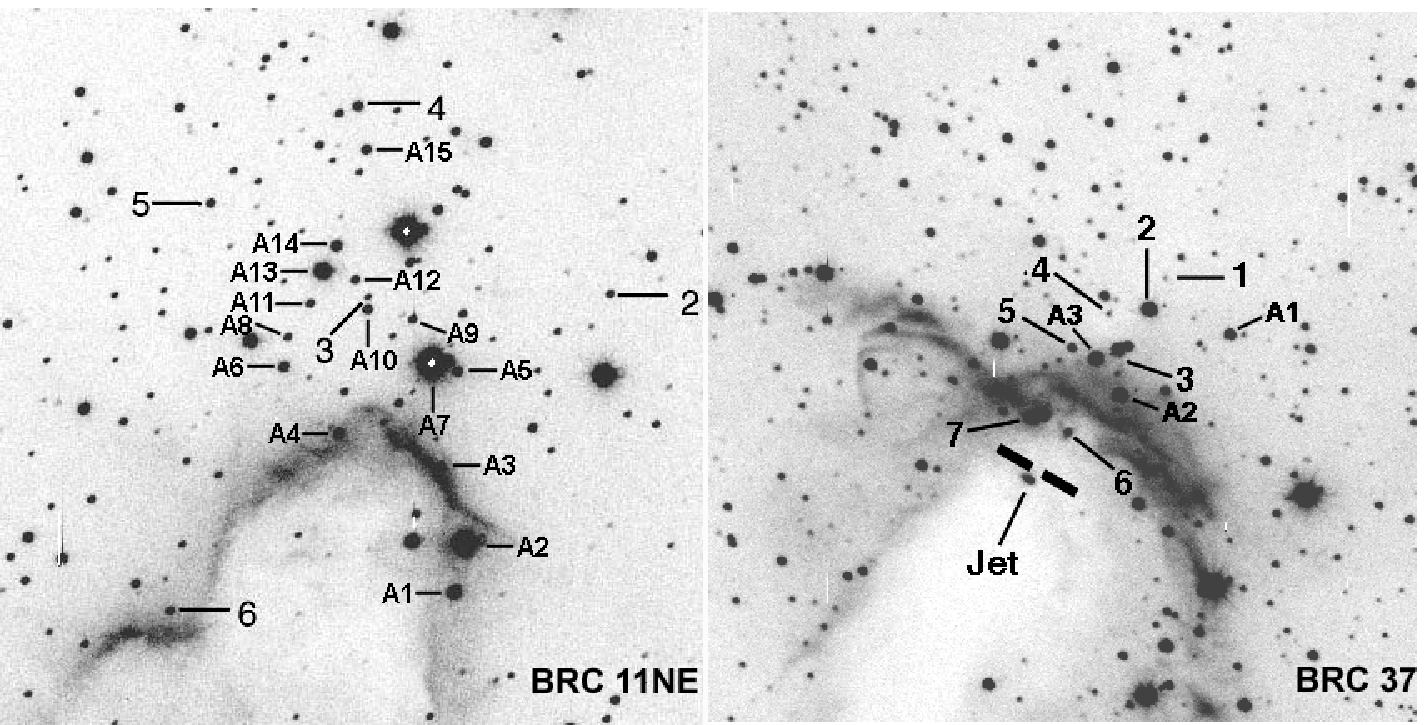}
    %%%\FigureFile(width,height){filename}
  \end{center}
  \caption{
  
Finding charts for stars in BRC 11NE and BRC 37. Stars with prefix "A" are non-H$\alpha$ emission stars, and those without the prefix are H$\alpha$ emitters of \citet{ogu02}. They are reproduced from their figures 1f and 1p, and are $2.5'$ by $2.5'$ wide each with north up and east to the left. In BRC 37 "jet" is part of the Herbig-Haro object HH 588 (see, \cite{ogu02}) and the pair of thick tick marks indicate the nominal IRAS position. BRC 11NE does not harbor any IRAS point sources.
 
}\label{fig:sample}
\end{figure}

\section{Results}

The photometric results for each star in each BRC aggregate are given in table 2. The second and third columns give the extinction-corrected {\it V} magnitude and {\it V-I$_{c}$\/} color, respectively. The last column gives the location of the star with respect to the bright rim. Note that "outside/inside BR" (BR stands for bright rim) are the inside/outside of the H\emissiontype{II} region. We cannot tell the accuracies of the values in Table 2, since the amounts of reddening corrections depend on several factors whose reliability is difficult to quantify. For each BRC a {\it V}$_{0}$/({\it V-I$_{c}$\/})$_{0}$ CMD  was constructed, where overlaid are the isochrones of \citet{sie00} with {\it Z\/} = 0.02 and no overshooting, which are corrected for the distance given in table 2. These distances are taken from the SFO Catalog (\cite{sug91}). The age of each star has been determined according to these isochrones. In view of the limited page space we show as an example the {\it V}$_{0}$/({\it V-I$_{c}$\/})$_{0}$ CMD for BRC 14 in figure 2. There are many stars which lie above the youngest isochrone of 1 Myr and their ages are estimated by extrapolation. The resultant ages appear to be reasonable, ranging from 0.1 to a few Myr (with only a few exceptions), which compare well with TTSs' lifetime of a few tenths to a few Myr. From this we would say that all the stars in our sample are probably aggregate members with practically no foreground/background stars mixed-in. The mass of the stars ranges from $\sim$0.3 to $\sim$2 \MO (for the 1 Myr isochrone), which is again reasonable. With these ages and masses the TTS nature of the H$\alpha$ emission stars is almost confirmed. {\it V\/}$_{0}$/({\it B-V\/})$_{0}$ CMDs yield similar results.

In each aggregate the locations of the stars with respect to the bright rim are grouped into two or three (only in the case of BRC 14) regions. The grouping is chosen so that there are at least four stars in each region with the exception of BRC 37 where we have altogether only six stars. The mean age and {\it A$_{V}$\/} of the stars in each region are summarized in table 3. It shows that the mean age of the stars inside or on the bright rim is {\it  always} younger than that outside it, although the scatter is very large. The age difference is particularly significant in BRCs 12 and 14 where we have stars inside the bright rim. Moreover in BRC 14 the mean age constantly decreases from outside to inside the bright rim. Therefore our results appear to confirm the {\it S\/$^4$F\/} hypothesis. 

From table 3 we see that stars inside the bright rim have larger {\it A$_{V}$\/} values than those outside it in the case of BRCs 12 and 14, although the scatters are large again. This confirms the result of \citet{mat06} for BRC 14 (but with a smaller difference between the regions in our case). It is natural in view of the above general trend of the stellar ages. In the case of BRCs 11NE and 37, however, no difference is found between the two regions.

%%%%%%%%%%%%%%%%%%%%%%%%%%%%%%%%%%%%%%%
\begin{table}
  \caption{Results of dereddening and age estimation}\label{tab:second}
  \begin{center}
    \begin{tabular}{lllllll}
     \hline\hline   
     Star &  {\it V\/}$_{0}$ & ({\it V-I$_{c}$\/})$_{0}$ &  ({\it B-V\/})$_{0}$ & {\it A$_{V}$\/} & Age & Location \\         
   ~ & ~ & ~ & ~ & (mag) & (Myr) & ~ \\        
  \hline\hline  					
%   BRC 11NE & ({\it d\/} $\sim$ 1.9 kpc) & ~ & ~ & ~ & ~ & ~ \\
   BRC 11NE & (1.9 kpc) & ~ & ~ & ~ & ~ & ~ \\
  \hline
  H$\alpha$-6 & 17.99 & 1.84 & 1.23 & 1.46 & 1.7 & on BR \\ 
        A1 & 15.45 & 1.57 & 1.22 & 1.41 & 0.3 & " \\
        A2 & 14.56 & 1,57 & 1.41 & 0.00 & 0.1 & " \\
        A3 & 17.44 & 2.00 & 1.40 & 0.71 & 0.7 & " \\
        A4 & 16.25 & 1.34 & 0.80 & 2.26 & 1.5 & " \\
        A5 & 17.91 & 1.75 & -0.21 & 0.69 & 1.9	 & outside BR \\
        A6 & 17.01 & 1.67 & 1.29 & 0.78 & 1.0 & " \\
        A7 & 13.13 & 0.85 & 0.53 & 0.00 & 1.0 & " \\
        A8 & 19.13 & 2.24	 & 1.65 & 0.78 & 1.9 & " \\
        A9 & 17.27 & 1.61 & -0.65 & 2.16 & 1.8 & " \\
        A10 & 15.56 & 1.58 & 1.06 & 3.09 & 0.3 & " \\
        A11 & 17.92 & 1.8 & 1.32 & 1.55 & 1.8 & " \\
        A12 & 16.84 & 1.45 & 1.11 & 2.29 & 1.9 & " \\
        A13 & 15.52 & 1.54 & 1.30 & 0.00 & 0.3 & " \\
        A14 & 16.26 & 1.63 & 1.21 & 1.42 & 0.6 & " \\
        H$\alpha$-5 & 18.32 & 2.08 & 1.52 & 0.73 & 1.4 & " \\
        A15 & 17.24 & 1.82 & 1.35 & 1.38 & 0.8 & " \\
        H$\alpha$-4 & 18.09 & 2.05 & 1.37 & 0.61 & 1.2 & " \\
    \hline\hline   
     BRC 12 & (1.9 kpc) & ~ & ~ & ~ & ~ & ~ \\ 
    \hline 
     H$\alpha$-1 & 16.19 & 1.61 & 1.06 & 2.61 & 0.6 & outside BR \\
      H$\alpha$-5 & 17.88 & 1.52 & 1.09 & 0.00 & 6.0 & " \\
      H$\alpha$-10 & 19.79 & 2.22 & 0.97 & 0.00 & 4.0 & " \\
      H$\alpha$-17 & 17.96 & 1.88 & 1.29 & 1.06 & 1.6 & " \\
      H$\alpha$-19 & 16.34 & 1.70 & 1.09 & 0.94 & 0.35 & " \\
      H$\alpha$-09 & 16.34 & 1.95 & 1.18 & 1.62 & 0.17 & on BR \\
      H$\alpha$-11 & 17.18 & 2.10 & ~ & 2.36 & 0.35 & " \\
      H$\alpha$-14 & 14.81 & 1.57 & 0.99 & 1.73 & 0.1 & " \\
      H$\alpha$-16 & 16.9 & 1.83 & 1.13 & 1.41 & 0.4 & " \\
      H$\alpha$-20 & 15.24 & 1.29 & 0.64 & 3.54 & 0.5 & inside BR \\   
   \hline\hline   
     BRC 14 & (1.9 kpc) & ~ & ~ & ~ & ~ & ~ \\ 
    \hline 
     H$\alpha$-5 & 16.99 & 1.84 & 1.50 & 2.18 & 0.5 & outside BR \\
     H$\alpha$-12 & 16.51 & 1.46 & 1.35 & 3.40 & 1.3 & " \\
      H$\alpha$-16 & 17.85 & 1.85 & ~ & 3.37 & 1.5 & " \\
      H$\alpha$-18 & 19.07 & 2.11 & ~ & 2.42 & 2.5 & " \\
      H$\alpha$-20 & 17.29 & 1.62 & ~ & 3.14 & 1.7 & " \\
      H$\alpha$-23 & 16.94 & 1.39 & ~ & 3.23 & 3.0 & " \\
      H$\alpha$-24 & 18.93 & 2.55 & ~ & 2.14 & 1.0 & " \\
      H$\alpha$-32 & 15.2 & 1.36 & 1.15 & 3.22 & 0.4 & on BR \\
      H$\alpha$-33 & 17.87 & 1.77 & ~ & 2.87 & 1.9 & " \\
      H$\alpha$-35 & 16.76 & 1.82 & 1.47 & 2.70 & 0.4 & " \\
      H$\alpha$-31 & 15.6 & 1.39 & ~ & 3.99 & 0.5 & " \\
      H$\alpha$-34 & 16.46 & 1.39 & 0.88 & 4.17 & 1.7 & " \\
      H$\alpha$-38 & 17.68 & 1.97 & ~ & 2.65 & 0.8 & inside BR \\
      H$\alpha$-39 & 17.22 & 1.89 & ~ & 2.87 & 0.5 & " \\
      H$\alpha$-40 & 15.04 & 1.37 & ~ & 5.47 & 0.3 & " \\
      H$\alpha$-42 & 17.14 & 1.95 & ~ & 3.04 & 0.4 & " \\ 
    \hline\hline   
      BRC 37 & (0.75 kpc) & ~ & ~ & ~ & ~ & ~ \\ 
     \hline 
     A1 & 16.07 & 1.63 & 1.23 & 2.04 & 4.5 & outside BR \\
     H$\alpha$-2 & 14,10 & 1.49 & 1.12 & 2.27 & 0.6 & " \\
     H$\alpha$-3 & 14.13 & 1.46 & ~ & 3.71 & 0.7 & " \\
     A2 & 14.65 & 2.00 & 1.01 & 2.52 & 0.25 & on near BR \\
     A3 & 15.11 & 1.62 & 1.20 & 2.07 & 1.5 & "  \\
     H$\alpha$-7 & 14.39 & 1.49 & 0.68 & 3.21 & 1.0 & on BR \\    
     \hline
     \end{tabular}
  \end{center}
\end{table}

%%%%%%%%%%%%%%%%%%%%%%%%%%%%%%%%%%%%%%%

\begin{figure}
  \begin{center}
    \FigureFile(80mm,57mm){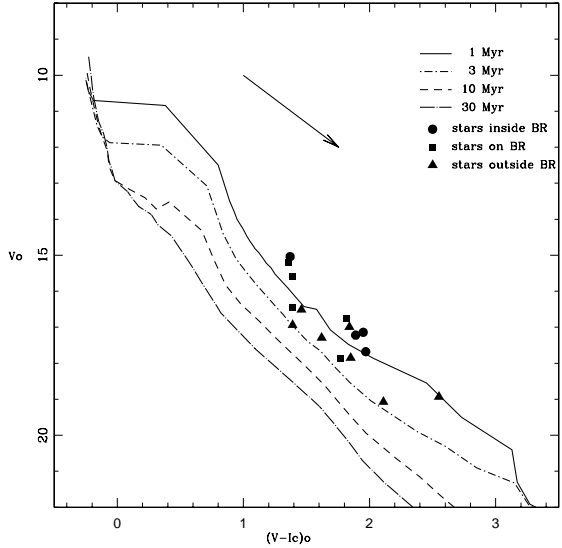}
    %%%\FigureFile(width,height){filename}
  \end{center}
  \caption{  {\it V\/}$_{0}$/({\it V-I$_{c}$\/})$_{0}$ CMD for the BRC 14 aggregate. Stars, which are all H$\alpha$ emitters in this case, are plotted with different symbols according to their locations with respect to the bright rim. The isochrones of \citet{sie00} with {\it Z\/} = 0.02 and no overshooting are drawn with the correction for the distance of 1.9 kpc. The arrow indicates the reddening vector.      }
\end{figure}

%%%%%%%%%%%%%%%%%%%%%%%%%%%%%%%%%%%%%%%

\begin{table}
  \caption{Summary of results}\label{tab:third}
  \begin{center}
    \begin{tabular}{lllll}
     \hline\hline     
      BRC & Region & No. of  &  Mean age &  Mean {\it A$_{V}$\/} \\
      ~ & ~ & stars & $\pm$st dev (Myr) & $\pm$st dev (mag) \\
  \hline 					
      11NE & outside BR & 13 & 1.26 $\pm$0.60 & 1.21 $\pm$0.88 \\
       ~ & on BR & ~5 & 0.86 $\pm$0.71 & 1.17 $\pm$0.85 \\
    \hline 
     12 & outside BR & ~5 & 2.51 $\pm$2.43 & 0.93 $\pm$1.07 \\
      ~ & inside \& on BR & ~5 & 0.30 $\pm$0.17 & 2.13 $\pm$0.86 \\
     \hline 
     14 & outside BR & ~7 & 1.64 $\pm$0.86 & 2.84 $\pm$0.57 \\
      ~ & on BR & ~5 & 0.98 $\pm$0.75 & 3.39 $\pm$0.66 \\
      ~ & inside BR & ~4 & 0.50 $\pm$0.22 & 3.51 $\pm$1.32 \\
     \hline 
     37 & outside BR & ~3 & 1.93 $\pm$2.22 & 2.67 $\pm$0.91 \\
      ~ & on \& near BR & ~3 & 0.92 $\pm$0.63 & 2.60 $\pm$0.57 \\
     \hline
     \end{tabular}
  \end{center}
\end{table}

%%%%%%%%%%%%%%%%%%%%%%%%%%%%%%%%%%%%%%%

\section{Discussion and Conclusion}

\citet{lee05} argued that the presence of TTSs only near and/or outside the surface of BRCs would clearly indicate the evidence for triggered star formation, and in their study of BRCs in the Orion region they actually found TTSs only between the BRCs and the OB stars. In the present study we locate some TTSs inside the bright rim. But their {\it A$_{V}$\/} values show that they are not deeply embedded in the BRC. Therefore their apparent location may be due to a projection effect and they may actually be located near the front surface of the BRC, so in reality similar to the "on BR" stars. The difference between \citet{lee05}'s result and ours is probably due to the fact that our study is deeper and more small-scaled.

The regions in each BRC aggregate show big scatters of the ages of the member stars in spite of a clear trend of the mean ages. The reason for this is not clear, but we suspect that it is caused by the proper motions of newly born stars. \citet{jon79} showed that young stars in the Taurus molecular cloud have velocity dispersions of 1-2 km/s in one coordinate. If this value is applicable to BRC aggregates, the stars must have moved 2 pc or so (in projection) from their birth place in 1 Myr. Since the size of the BRC aggregates is $\sim$0.2 pc (BRC 37) to $\sim$2 pc (BRC 12), this is sufficiently large for some of the stars which formed 1 Myr ago to move into the (projected) inside of the present bright rim. We suspect that many stars in H\emissiontype{II} regions have thier origin in already dispersed BRC aggregates except for the case where a rich cluster exists.

The stars outside a BRC (i.e., inside the H\emissiontype{II} region) have similar ages irrespective of the (projected) distance from the bright rim. It is our position that they essentially formed there inside the BRC that has since retreated to its present position. Then the above fact suggests that they were formed in the (first) collapse phase of the evolution of BRCs (\cite{lef94}). Presumably BRCs may have a second (or, even third) collapse phase due to their clumpiness or as the exciting O star(s) evolve(s) and increase(s) its (their) UV photon emission, or somehow they give birth to stars after this phase (i.e., in the cometary phase); and the stars on or near the bright rim, which generally are younger, may correspond to such origin. It is interesting to note that among the aggregate regions in table 3, those having the youngest mean age (i.e,, inside/on BR of BRCs 12 and 14) show the smallest age scatter. 

The age estimate from CMDs depends on the adopted distance; if we use a larger distance we get younger ages. For some BRCs in the SFO Catalog the distances are not well determined. However,  since the results given in tables 2 and 3 are in a reasonable range and do not show any systematic differences between the BRCs, the distances adopted here are not in big error, presumably. Anyway, what is important in the present study is the relative ages among the stars in a BRC, and they are not changed by the error of the adopted distance. Another factor which affects the age estimation is duplicity of  stars. On a CMD an unresolved binary star can be placed at an elevated position (by up to 0.75 mag), which results in a younger age. Some of the stars which fall above the 1 Myr isochrone (see figure 2) might be such stars. However, unresolved binaries must not be limited to this location, but exist among all the stars more or less equally on the CMDs. Therefore our main conclusion remain unchanged by this effect. We therefore conclude that our photometry of the BRC aggregates confirms the {\it S\/$^4$F\/} hypothesis, although some problems remain.

%%%%%%%%%%%%%%%%%%%%%%%%%%%%
%\begin{longtable}{ll}
%  \caption{Sample of long tabular}\label{tab:LTsample}
%  \hline\hline
%&  name & value \\
%\endfirsthead
%  \hline\hline
%  name & value \\
%\endhead
%  \hline
%\endfoot
%  \hline
%\endlastfoot
%  aaaaa & bbbbb \\
%  ...... & ..... \\
%  yyyyy & zzzzz \\
%\end{longtable}
%%%%%%%%%%%%%%%%%%%%%%%%%%%%

\bigskip

We thank the TAC and staff of HCT for the time allotment and for the support in the observations, respectively. Our thanks are due also to I. Matsuyanagi and the SIRIUS team for providing the {\it JHK$_{s}$\/} data on BRC 14. We are grateful to the referee Dr. G.H. Herbig for his helpful comments.  
K.O. and A.K.P. acknowledge JSPS (Japan) and DST (India) for the finantial supports. Y.I. is supported by "The 21st Century COE Program: The Origin and Evolution of Planetary System" of MEXT.

\end{document}